\documentclass[
11pt,preprintnumbers]{article}

\usepackage{fullpage}
\usepackage[T1]{fontenc} 

\usepackage{amssymb,amsmath}

\usepackage{graphicx}

\usepackage{caption}

\usepackage{cite}

\allowbreak

\newcommand{\be}{\begin{eqnarray}}

\newcommand{\ee}{\end{eqnarray}}

\def\p{\partial}

\def\s{{\sigma}}

\def\p{\partial}

\begin{document} 
	
\begin{titlepage}
	\thispagestyle{empty}
	\begin{flushright}

		\hfill{DFPD-2017/TH/06}
	\end{flushright}

	\vspace{35pt}
	
	\begin{center}
	    { \Large{\bf Constrained superfields from inflation to reheating}}
		
		\vspace{50pt}

		{Ioannis Dalianis$^{1}$ and Fotis Farakos$^{2,3}$} 

		\vspace{25pt}

		{
			$^1${\it  Physics Division, National Technical University of Athens, 
			\\ \it 15780 Zografou Campus, Athens, Greece} 

		\vspace{15pt}

			$^2${\it  Dipartimento di Fisica e Astronomia ``Galileo Galilei''\\
			Universit\`a di Padova, Via Marzolo 8, 35131 Padova, Italy}

		\vspace{15pt}

			$^3${\it   INFN, Sezione di Padova \\
			Via Marzolo 8, 35131 Padova, Italy}

		}

		\vspace{40pt}

		{ABSTRACT}
	\end{center}
	
	\vspace{10pt} 
\noindent 
We construct effective supergravity theories from customized constrained superfields 
which provide a setup consistent both for the description of inflation and the subsequent reheating processes. 
These theories contain the minimum degrees of freedom in the bosonic sector required for single-field inflation.

\bigskip

\end{titlepage}

\baselineskip 6 mm


\section{Introduction}

Constrained superfields \cite{Rocek:1978nb,Lind,Casalbuoni:1988xh,Komargodski:2009rz,Cribiori:2017ngp,DallAgata:2016syy} 
provide an elegant, albeit effective, method for embedding inflation in 
supergravity \cite{Antoniadis:2014oya,Ferrara:2014kva,Kallosh:2014via,DallAgata:2014qsj,
Kahn:2015mla,Ferrara:2015tyn,Carrasco:2015iij,DallAgata:2015zxp,Kallosh:2016ndd,McDonough:2016der}. 
Requiring the effective theory 
to be valid from inflation down to the present de Sitter phase of the universe 
results in important restrictions. 
For example, various consistency conditions were already pointed out in \cite{DallAgata:2014qsj,Dudas:2016eej} for {\it sgoldstino-less} models. 
The utility of the constrained superfields extends to aspects of the supergravity cosmology besides inflation, see e.g. \cite{Hasegawa:2017hgd, Benakli:2017whb} for recent works on gravitino cosmology. 
The description of the gravitino production in this context  is of particular interest and motivates this letter.

Gravitinos are cosmologically problematic when overabundant \cite{Ellis:1982yb, Nanopoulos:1983up, Ellis:1984eq}. 
The preheating and reheating stages \cite{Kofman:1997yn} are important sources of non-thermal gravitino production as has been demonstrated in notable works \cite{Maroto:1999ch, Giudice:1999fb, Lemoine:1999sc, Kallosh:1999jj, Giudice:1999yt, Giudice:1999am, Kallosh:2000ve, Nilles:2001ry, Nilles:2001fg, Greene:2002ku,  Kawasaki:2006hm, Ema:2016oxl}. 
The number density of the $\pm 1/2$ helicity gravitinos, which are identified as the goldstinos at energies much larger than the gravitino mass, is not suppressed by inverse powers of $M_\text{Pl}$ and it can be rather large \cite{ Kallosh:1999jj, Giudice:1999yt, Giudice:1999am, Kallosh:2000ve}. In the standard picture,
the goldstino is a linear combination of fermions, when several chiral superfields are present. 
Hence the composition of the helicity $\pm 1/2$ gravitino through the goldstino mixture varies due to the cosmic evolution. 
During the stage of the  inflaton oscillations, 
one assumes the main contribution to the helicity $\pm1/2$ gravitino to come from the 
fermionic superpartner of the inflaton, the inflatino, 
whereas at times after the reheating of the universe the longitudinal gravitino is given by the 
present vacuum golstino, 
which is the {\it true goldstino}. 
In \cite{Nilles:2001ry, Nilles:2001fg} it was understood that this change with time of the definition 
of the longitudinal gravitino renders the production of gravitinos during the preheating stage cosmologically harmless.

In the framework of constrained superfields the gravitino production during preheating and reheating was 
examined recently \cite{Hasegawa:2017hgd} for the inflatino-less 
models constructed in \cite{Kahn:2015mla,Ferrara:2015tyn,Carrasco:2015iij,DallAgata:2015zxp}. 
The inflatino-less models of \cite{Kahn:2015mla,Ferrara:2015tyn,Carrasco:2015iij,DallAgata:2015zxp} have the property 
to contain the minimal bosonic sector essential for single-field inflation, 
but due to the constraints that are imposed it is also the inflatino which is eliminated from the spectrum. 
It was mentioned in \cite{Hasegawa:2017hgd} that gravitinos may be found to be 
cosmologically overabundant when the specific constrained superfields are considered 
as the inflationary sector. 
Indeed, when one eliminates all the spin-1/2 fermions in the theory except for the goldstino, 
then 
the longitudinal gravitino modes will be identified with the 
gravitino in the vacuum and therefore naturally lead to large gravitino number densities. 
Also, the perturbative decay of the inflaton may produce an excessive gravitino yield if the inflaton 
contributes to the vacuum supersymmetry breaking. 
Triggered by this novel approach to the gravitino cosmology we revisit the non-linear realizations 
of supersymmetric cosmology after inflation 
in order to build effective field theories that achieve agreement with the existing standard supergravity results.

Aiming at minimal effective theories 
which can describe both inflation and reheating processes,
we introduce {\it customized} constrained superfields which contain 
minimum number of component fields. 
These models have a minimal bosonic sector which includes only the inflaton and the metric 
and deliver inflationary Lagrangians of the form  
\be
\label{introL}
e^{-1} {\cal L} = -\frac12 R - \frac12 \, \p_m \phi \, \p^m \phi - V(\phi) \, . 
\ee
The fermion sector is left unconstrained, 
therefore will include the {\it vacuum goldstino} $G_\alpha$, 
the gravitino $\psi_m^\alpha$, 
and the fermions of the inflationary sector: {\it inflatino} $\chi_\alpha$, and {\it stabilino} $\lambda_\alpha$ when there is a stabilizer superfield. 
The number of constrained superfields in the models we construct is {\it not} essentially minimal, 
but this gives us the required flexibility to have a consistent effective description during the various stages of the evolution of the universe. 
In particular we make use of a constrained 
inflaton superfield $\Phi$, 
a constrained stabilizer superfield $S$, 
and the nilpotent chiral superfield $X$ which will break supersymmetry in the vacuum.

The organization of the article is the following.
In the second section we revisit the inflatino-less models and discuss the origin of the gravitino overproduction cosmological problem during the non-thermal (p)reheating stage. 
In the third section we present the customized constrained superfields adequate to describe the Polonyi sector, 
the inflaton sector and the stabilizer sector. 
In the fourth section we present two working models, 
described in subsections (4.1) and (4.2),
that realize our proposal and in this context we review some relevant (p)reheating results.  In section five we conclude.

\section{Inflatino-less models and the gravitino issue}

The first supergravity inflationary models that contained the minimal bosonic sector (inflaton and gravitation) 
were constructed by two constrained superfields \cite{Kahn:2015mla,Ferrara:2015tyn,Carrasco:2015iij,DallAgata:2015zxp}: 
the chiral $X$ satisfying $X^2=0$ and a second chiral superfield ${\cal A}$ satisfying \cite{Komargodski:2009rz} 
\be
\label{XAXA}
X {\cal A} = X \overline{\cal A} \, . 
\ee
The effect of the constraint \eqref{XAXA} is to eliminate all the component fields of the superfield ${\cal A}$ 
leaving as independent only a real scalar residing in the lowest component field which is identified with the 
inflaton $\phi$. Explicitly one finds: ${\cal A}| = \phi + \text{terms with goldstino}$. 
The crucial side-effect of the constraint \eqref{XAXA} is that 
the {\it inflatino} is also eliminated which would otherwise sit in the fermion component of the superspace expansion of ${\cal A}$.

After the end of inflation, gravitino production generically takes place during both phases of the reheating process: 
initially during preheating where the inflaton \emph{oscillates} and excites the other fields in the theory, 
and subsequently during the perturbative \emph{decay} of the inflaton. 
The gravitino production is cosmologically problematic when the corresponding yield  $Y_{3/2}\equiv n_{3/2}/s$, where $s$ the entropy density and  $n_{3/2}$ the gravitino number density, violates either the cold dark matter density constraint \cite{Ade:2015xua}
\begin{equation} \label{c1}
\Omega_{3/2}h^2 \, \sim  \, 10^8 \, \left( \frac{m_{3/2}}{\text{GeV}}\right)\, Y_{3/2}\, h^2\, \lesssim \, 0.12 \, , 
\end{equation}
for stable gravitinos, or the Big-Bang nucleosynthesis constraints
\begin{equation} \label{c2}
 Y_{3/2} \lesssim 10^{-16} -10^{-13} \,  , 
\end{equation}
for unstable gravitinos with mass $m_{3/2}\lesssim 10^5$ GeV (see e.g. \cite{Kawasaki:2006hm, Ellis:2015jpg} 
for recent work and references therein). 
The $Y_{3/2}$ may be additionally constrained depending on the details of the LSP dark matter production mechanism.

Under general assumptions, 
it is known that during the preheating phase non-thermal production of longitudinal gravitinos takes place \cite{Kallosh:1999jj, Giudice:1999am}. 
This issue came with a simple resolution \cite{Nilles:2001ry}. Since both the kinetic 
and potential energies contribute to the supersymmetry breaking \cite{Giudice:1999am}, the gravitino longitudinal component is dominated by the inflatino during inflation and preheating, 
and therefore the gravitino longitudinal components during these stages are not true goldstinos. 
They are rather harmless inflatinos which are produced during preheating and do not correspond to gravitinos in the vacuum. 
Essentially, the source of supersymmetry breaking during inflation and preheating has to be 
different from the source of supersymmetry breaking in the vacuum.

In the recent work \cite{Hasegawa:2017hgd}, it has been pointed out that the 
inflatino-less models \cite{Kahn:2015mla,Ferrara:2015tyn,Carrasco:2015iij,DallAgata:2015zxp} 
generically suffer from a gravitino overproduction problem after inflation. 
Indeed, in contrast to the standard supergravity situation where inflatinos instead of gravitinos
are produced during preheating, 
in this setup this is not possible. 
The reason is that when using the constraint \eqref{XAXA}, 
the inflatino is eliminated in terms of the goldstino. 
Therefore the longitudinal component of the gravitino is always aligned with the goldstino $G_\alpha$ 
during the cosmological phases from inflation to reheating and to the current 
de Sitter vacuum \cite{Dudas:2015eha,Bergshoeff:2015tra,Hasegawa:2015bza,Bandos:2015xnf,Bandos:2016xyu,Cribiori:2016qif}. 
This means that any longitudinal gravitino production is essentially a {\it true goldstino} production 
and might easily lead to cosmological problems exactly as has been noted in the early literature \cite{Kallosh:1999jj, Giudice:1999am}.

We interpret the findings of \cite{Hasegawa:2017hgd} as an indication 
that the constrained superfield ${\cal A}$ 
might not 
always describe the inflaton in an appropriate way. 
In this work we construct effective theories 
with the appropriate inflaton superfield 
which can describe the interactions of the inflatino and still maintain a minimal bosonic sector. 
In this way minimal inflationary models with constrained superfields can produce the standard  
supergravity results regarding the gravitino production \cite{Nilles:2001ry}.

\section{Customized constrained superfields}

In this section we wish to introduce the constrained superfields which contain the minimal number of independent fields, 
while still serving the purpose they are introduced for in standard supergravity. 
The supergravity Lagrangians we will consider in this work are always built by the standard superspace couplings  
\begin{equation}
\label{L1}
{\cal L} = \int d^2 \Theta \, 2 {\cal E}\, \left[ \frac38 \left( \overline {\cal D}^2 - 8 {\cal R} \right) \, \text{e}^{-K/3} + W \right] + c.c. \, , 
\end{equation} 
where we have set $M_\text{Pl}=1$.  
The expressions for the chiral density $2 {\cal E}$ and the curvature superfield ${\cal R}$ can be found in \cite{Wess:1992cp}, 
where also the component form expression for \eqref{L1} is given.

{\it The Polonyi superfield.} 
The Polonyi model provides the simplest realization of supersymmetry breaking in supergravity. 
In this setup one introduces a K\"ahler potential and superpotential of the form 
\be
K = X \overline X  \ , \quad W = f_0 X + W_0 \, . 
\ee
By appropriately choosing the parameters $f_0$ and $W_0$ in the superpotential and 
introducing higher order corrections in the K\"ahler potential one obtains a vacuum with a small enough cosmological constant and 
a heavy scalar which essentially decouples. 
The effective description of this setup is provided by the nilpotent constraint superfield $X$ 
which has found its way into various applications for supergravity cosmology \cite{Ferrara:2016ajl}. 
Indeed, 
the decoupling of the heavy Polonyi scalar corresponds to the constraint 
\be
\label{X2}
X^2 =0 \, , 
\ee
which is solved for $X = \frac{G^2}{2 F^X} + \sqrt 2 \Theta G + \Theta^2 F^X$, 
under the strict requirement 
\be
\label{FXvac}
\langle F^X \rangle \ne  0 \, .  
\ee
This setup is tailor-made to describe the current de Sitter vacuum of the universe. 
The pure theories were constructed in \cite{Dudas:2015eha,Bergshoeff:2015tra,Hasegawa:2015bza}, 
and the vacuum energy and gravitino mass are given by 
\be
\Lambda = f_0^2 - 3 \, W_0^2 \ , \quad m_{3/2} = W_0 \, . 
\ee 
Throughout this work we will always require that the nilpotent Polonyi 
field $X$ is the sole source responsible for supersymmetry breaking at the vacuum, 
therefore the $G_\alpha$ is the true vacuum goldstino. 
Any production of $G_\alpha$ corresponds to production of gravitino. 
Notice that since in the vacuum $F^X$ is the only auxiliary field which will get a non-vanishing vacuum expectation value 
it is in fact singled out to be the one belonging to the superfield satisfying \eqref{X2}. 
Moreover, 
since the goldstino is a pure gauge and will be absorbed by the gravitino, 
it will be convenient to express our results directly in the gauge 
\be
\label{G0}
G_\alpha = 0 \, . 
\ee
In this gauge $X$ becomes $X|_{G=0} =  \Theta^2 F^X$.

{\it The inflaton superfield.} 
We will now present a new class of constrained superfields  which are adequate for hosting the inflaton in their lowest component. 
The purpose of this constraint is to eliminate \emph{only} 
the real scalar $b$ residing in the imaginary part of $\Phi|$. 
Indeed, once we have the nilpotent $X$ superfield, 
the method to construct customized superfields was outlined in \cite{DallAgata:2016syy}. 
Here we will use this method to construct the inflaton superfield. 
To this end we simply have to impose the constraint on $\Phi$ 
\be
\label{inflaton}
X \overline X \left( \Phi - \overline \Phi \right) = 0 \, . 
\ee
To solve this constraint, 
one first brings it to the form 
\begin{equation}
\begin{aligned}
\left( \Phi - \overline \Phi \right) = \, & 
2 \frac{\overline{\cal D}_{\dot\alpha}\overline X}{\overline{\cal D}^2 \overline X} \, \overline{\cal D}^{\dot \alpha} \overline \Phi  
+ \overline X \frac{\overline{\cal D}^2 \overline \Phi }{\overline{\cal D}^2\overline X} 
- 2\frac{{\cal D^\alpha}X}{{\cal D}^2 X\,\overline{\cal D}^2\overline X} {\cal D}_\alpha \overline{\cal D}^2 
\left[ \overline X \left( \Phi - \overline \Phi \right) \right]  
\\
& 
-\frac{X}{{\cal D}^2 X \overline{\cal D}^2 \overline X}{\cal D}^2 \overline{\cal D}^2 \left[ \overline X \left( \Phi - \overline \Phi \right) \right] \, , 
\end{aligned}
\end{equation}
and projects to $\theta=0$ to find the component field expression. 
Then by performing recursive steps the real scalar component field $b$ will be eliminated from the spectrum, 
in terms of the other component fields of $\Phi$ and $X$. 
The leading terms in the $G$ expansion of $b = \left( \Phi - \overline \Phi \right)|/2i$ are 
\be
\label{bEXPANSION}
b = - i \frac{G \chi}{\sqrt 2 \, F} 
+ i \frac{\overline G \overline \chi}{\sqrt 2 \overline F} 
+ i \frac{G^2}{4 F^2} F^\Phi - i \frac{\overline G^2}{4 \overline F^2} \overline F^\Phi  
- \frac{1}{2 F \overline F} \left( G \sigma^c \overline G \right) e_c^m \p_m \phi + \cdots 
\ee
where the $\cdots$ refer to terms with at least three goldstini. 
In \eqref{bEXPANSION} we used the component field definitions: $\chi_\alpha={\cal D}_\alpha\Phi|/2$ 
and 
$F^\Phi = - {\cal D}^2 \Phi|/4$. 
Note that \eqref{bEXPANSION} is the solution in curved superspace, 
but the gravitino and the auxiliary fields of the supergravity sector do not enter at this order. 
During inflation, 
the effect of \eqref{inflaton} on $\Phi$ is more transparent once  we write the inflaton superfield in the simplifying gauge \eqref{G0}. 
Indeed in this gauge we will have 
\be
\Phi|_{G=0} = \phi + 2 \, \Theta \chi + \Theta^2 F^\Phi \, ,  
\ee
where $\phi$ is a real scalar (the inflaton), $\chi$ is the inflatino, 
and $F^\Phi$ is the auxiliary field of the inflaton multiplet. 
The constraint \eqref{inflaton} can be imposed with a Lagrange multiplier in the spirit of \cite{Ferrara:2016een}.

The importance of introducing the inflatino and the auxiliary field of the inflaton multiplet has been stressed in the previous section. 
We recall that 
if the inflatino is eliminated then the longitudinal gravitino component production during preheating 
might lead to an overestimate of the gravitino number density. 
If the inflatino survives in the effective theory then it is in fact the inflatino which is dominantly produced during preheating \cite{Nilles:2001ry}. 
We stress at this point that  by just introducing the inflatino in the effective theory 
one does {\itshape not} automatically have a viable gravitino cosmology. 
Rather now the gravitino cosmology has the same status as it is in standard supergravity
(see e.g. \cite{Kawasaki:2006hm, Ellis:2015jpg}). 
In the inflatino-less models  
this is not the case and one has to find new means of addressing this issue \cite{Hasegawa:2017hgd}.

{\it The stabilizer superfield.} The third ingredient for the inflationary model building in standard supergravity is the stabilizer superfield $S$. 
This superfield has a dual role. 
First it has specific couplings which allow to construct a large class of inflationary potentials for the inflaton superfield \cite{Kallosh:2010xz}. 
Second, 
if it is the primary source of supersymmetry breaking during inflation it is used to strongly stabilize all scalars (during inflation), 
except for the inflaton. 
The important component field of the stabilizer superfield is its auxiliary field $F^S$ and it should survive in the effective theory. 
Moreover, we wish to keep also the fermionic field of the stabilizer superfield, 
the {\it stabilino}. 
The reason is simple: during the preheating stage one has $F^S \neq 0$ 
therefore 
if the fermion of the stabilizer is eliminated from the spectrum its contribution to the 
longitudinal gravitino will be misinterpreted as 
true vacuum goldstinos $G_\alpha$. 
On the other hand, 
the scalar of the stabilizer superfield can be very heavy and decouple. 
Therefore the appropriate constraint that should be imposed to describe the stabilizer superfield is 
\be
X \overline X \, S = 0 \, . 
\ee
The solution to this constraint and minimal models can be found in \cite{Brignole:1997pe,Komargodski:2009rz,DallAgata:2015pdd}. 
For our purposes it suffices to present $S$ in the gauge \eqref{G0}, 
where it has the form 
\be
S |_{G=0} = \sqrt 2 \Theta \, \lambda + \Theta^2 F^S \, . 
\ee
The stabilizer will essentially couple to the inflaton superfield in the superpotential via terms of the form $S \, \Sigma(\Phi)$, 
which will reproduce the contribution to the scalar potential 
\be
|F^S|^2 \sim |\Sigma (\phi)|^2 \, . 
\ee

In the subsequent section we will construct effective theories from these superfields 
which can describe single field inflation, the reheating phase and the current de Sitter phase of our universe. 
We will require that 
\be
\label{FFvac}
\langle F^\Phi \rangle |_\text{vacuum} = 0 \ , \quad  \langle F^S \rangle |_\text{vacuum} = 0 \, , 
\ee
such that any inflatino or stabilino production will not correspond to true vacuum goldstino.

\section{Effective theories}

\subsection{Models with the inflaton and the Polonyi superfields}

In this section we will study models which contain only the inflaton superfield $\Phi$ 
and the Polonyi superfield $X$. 
We will not write the most general coupling, 
but we will rather construct simple inflationary models which however capture 
most of the possibilities for single field inflation. 
Since we present these supergravity effective theories here for the first time, we will study them in more detail. 
The K\"ahler potential and the superpotential  we insert in \eqref{L1} are 
\be
\label{KW11}
\begin{split}
K =& X \overline X  - \frac14 (\Phi - \overline \Phi )^2 \, , 
\\
W =&  f_0 \,  X + g(\Phi) \,  , 
\end{split}
\ee
where  $f_0$ is a real constant. 
The K\"ahler potential has the shift symmetry $\Phi \rightarrow \Phi + c$ (for $c$ real constant) as 
proposed by \cite{Kawasaki:2000yn} to address the supergravity $\eta$-problem. 
Notice that the $X$ and $\Phi$ sectors are separated and will therefore interact mostly gravitationally. 
We could have chosen instead of $f_0$ a function $f(\Phi)$, but we prefer to study a minimal setup here. 
In the $G_\alpha=0$ gauge the full theory now reads 
\be 
\label{XFlagr22}
\begin{split}
e^{-1} {\cal L} = & -\frac12 R 
+ \frac{1}{2} \epsilon^{klmn} \left( \overline \psi_k \overline \s_l {\cal D}_m \psi_n - \psi_k \s_l {\cal D}_m \overline \psi_n \right) 
-  i \, \overline \chi \overline \s^m {\cal D}_m \chi 
\\
&- \frac12 \partial_m \phi \, \chi \s^m \overline \s^n \psi_m 
- \frac12 \partial_n \phi \, \overline \chi \overline \s^m \s^n \overline \psi_m 
+ \frac14 \left[ i \epsilon^{klmn} \psi_k \sigma_l \overline \psi_m 
+ \psi_m \sigma^n \overline \psi^m \right] \chi \sigma_n \overline \chi 
\\
& 
- \frac{1}{8} \chi^2 \overline \chi^2 
- i g_\phi \chi \s^a \overline \psi_a 
- i \overline g_\phi \overline \chi \overline \s^a \psi_a  
-  \left( g_{\phi \phi} - \frac12 g \right) \chi^2 
-  \left( \overline g_{\phi \phi} - \frac12 \overline g \right) \overline \chi^2 
\\
&
- \left( g(\phi)  \overline \psi_a \overline \s^{ab} \overline \psi_b 
+ \overline g(\phi)  \psi_a \s^{ab} \psi_b \right)  
- \frac12 \p^m \phi \, \p_m  \phi 
- V(\phi) \, . 
\end{split}
\ee 
The scalar potential in \eqref{XFlagr22} now has the form expected from standard supergravity 
\be
\label{V11}
V(\phi) = f_0^2 + 2 \Big{|} \frac{\p g(\phi)}{\p \phi} \Big{|}^2 - 3 |g(\phi)|^2 \, . 
\ee
The term $+ 2 \Big{|} \frac{\p g(\phi)}{\p \varphi} \Big{|}^2$ in particular comes exactly from integrating out the auxiliary field $F^\Phi$. 
Notice that the bosonic sector in \eqref{XFlagr22} is \emph{minimal}: 
it contains only the inflaton and gravitation. 
We will show in a moment how to construct simple models for inflation from  the expression \eqref{V11}.

The estimation of the gravitino abundance during preheating goes along the lines of the standard supergravity. 
The gravitino is not overproduced once a hierarchy between the inflationary scale and the supersymmetry breaking due to the Polonyi superfield is invoked. 
The inflationary scale is generically characterized by the inflaton mass in the vacuum $m_\phi$, 
while the vacuum supersymmetry breaking by the gravitino mass. 
In this case one requires \cite{Nilles:2001ry} 
\be
\label{mmvac}
m_\phi|_\text{vacuum} \gg m_{3/2}|_\text{vacuum} \, , 
\ee
and the conditions \eqref{FXvac} and \eqref{FFvac} to hold in the vacuum. 
Therefore in these models, during preheating supersymmetry breaking is dominated by the inflaton energy density 
and the longitudinal component of the gravitino should be identified with the inflatino $\chi_\alpha$ 
instead of the {\it true goldstino} $G_\alpha$ 
\cite{Nilles:2001ry}.

Let us present a simple realization of inflation in these models. 
In fact our construction here is very similar to the {\it single superfield} inflationary models \cite{Ketov:2014qha,Ketov:2016gej}, 
and we require that the inflationary sector does not break supersymmetry in the vacuum. 
Consider the function entering the superpotential \eqref{KW11} to have the form 
\be
g = \text{e}^{2i \, \Phi} h(\Phi) + h_0 \, , 
\ee
where $h_0$ is a real constant. 
The scalar potential then reads 
\be
V = 2 h_\phi^2 + 5 h^2 +(f_0^2-3h_0^2) -6 \, h h_0 \,  \text{cos}(2 \phi ) \, , 
\ee
which can drive inflation by appropriately choosing the function $h(\phi)$. 
We also require to have a function $h$ such that 
\be
\label{vacvac}
\langle g_\phi \rangle = 0 = \langle h_\phi \rangle = \langle h \rangle  \, , 
\ee 
which then automatically satisfies \eqref{FFvac}. 
The vacuum energy after the end of inflation will be $\Lambda_0 = f_0^2 - 3 h_0^2$ 
which we assume to be very small.

An example for $h$ is 
\be
h = \sqrt{V_0} \left[1- (1 + \phi ) \text{e}^{-\phi} \right] \, , 
\ee
with $\sqrt{V_0} \gg h_0$. 
The scalar potential becomes effectively 
\be
\label{SCSC}
V \sim V_0 \left\{ 2 \phi^2 \text{e}^{-2 \phi} + 5 \left( 1- \text{e}^{-\phi}  - \phi \, \text{e}^{-\phi}  \right)^2 \right\} \, , 
\ee
which for large $\phi$ values approaches the constant value $5 \times V_0$ and drives inflation with a plateau potential. 
In particular this model predicts for the tensor to scalar ratio: $r \sim 3 \times 10^{-3}$ 
and for the spectral index: $1 -n_s \sim 0.035$, 
which are consistent with the latest data released by the Planck collaboration \cite{Ade:2015lrj}. 
Moreover one can verify that the vacuum is at $\langle \phi \rangle = 0$, 
which then implies \eqref{vacvac}. 
Notice that during inflation $m_{3/2} > H^2$, 
therefore these models are within the perturbative unitarity bound discussed in \cite{DallAgata:2014qsj}. 
For the inflaton and gravitino masses in the vacuum we have 
\be
m_\phi|_\text{vacuum} \sim \sqrt{V_0} \ , \quad m_{3/2}|_\text{vacuum} = h_0 \, , 
\ee
which realize \eqref{mmvac}.

\subsection{Models including the stabilizer superfield} 

In this section we will construct inflationary models which on top of the inflaton superfield $\Phi$ and the Polonyi superfield $X$ 
also include the stabilizer superfield $S$. 
The fact that there is now the stabilizer in the effective theory gives much more freedom in the inflationary model building. 
We can construct generic models by using the K\"ahler potential 
\be
\label{K22}
K = X \overline X + S \overline S  - \frac14 (\Phi - \overline \Phi )^2 \, , 
\ee
and the superpotential 
\be
\label{W22}
W = f_0 \,  X + \Sigma(\Phi) \, S + g(\Phi) + g_0 \, ,  
\ee
where $g_0$ is a complex constant. 
Once we insert \eqref{K22} and \eqref{W22} into \eqref{L1} 
we find a theory with the minimal bosonic sector required for single field inflation, 
therefore the Lagrangian is \eqref{introL}, 
where the scalar potential reads 
\be
\label{29}
V(\phi) = f_0^2 + |\Sigma(\phi)|^2  +  2 \Big{|} \frac{\p g(\phi)}{\p \phi} \Big{|}^2 - 3 |g(\phi) + g_0|^2 \, . 
\ee
The gravitino mass is given by 
\be
\label{30}
m_{3/2} = |g(\phi) + g_0| \, . 
\ee
Now the condition \eqref{mmvac} can be easily satisfied because the gravitino mass \eqref{30} 
can be disentangled from the inflationary potential \eqref{29}. 
We also require that in the vacuum $\langle \Sigma(\phi) \rangle = \langle g_\phi(\phi) \rangle =0$, 
such that \eqref{FFvac} are satisfied.

Simple models can be constructed by mimicking \cite{DallAgata:2014qsj} and setting 
\be
\Sigma(\Phi) = \sqrt 3 \, g(\Phi) \, , 
\ee
where now $g(\phi)= \left(g(\phi)\right)^*$ but $g_0 = - g_0^*$ in \eqref{W22}, which brings the scalar potential to the form 
\be
V(\phi) = \left( f_0^2 - 3 |g_0|^2 \right)  +  2 \Big{|} \frac{\p g(\phi)}{\p \phi} \Big{|}^2 \, . 
\ee 
We can now construct various inflationary potentials $V_\text{infl.}(\phi)= 2 {\cal F}^2(\phi)$ by choosing  
\be
g(\Phi) = \int d\Phi \, {\cal F}(\Phi) \, , 
\ee
and fixing the integration constant such that $\langle g_\phi \rangle = \langle g \rangle =0$. 
For example, a realization of inflation with a plateau potential is achieved if we set 
\be
g(\Phi) =  \sqrt{V_0} \left(-\sqrt{\frac{3}{2}} + \Phi + \sqrt{\frac{3}{2}} \text{e}^{- \sqrt{\frac{2}{3}} \Phi} \right) \, , 
\ee
with $\sqrt{V_0} \gg |g_0|$. 
The scalar potential then reads 
\be
\label{STST}
V(\phi) = \left( f_0^2 - 3 |g_0|^2 \right)  + 2 V_0 \left(1 - \text{e}^{- \sqrt{\frac{2}{3}} \phi} \right)^2 \, .  
\ee
The scalar potential \eqref{STST} gives predictions for the $r$ and $1-n_s$ similar 
to the Starobinsky model of inflation. 
The vacuum of this theory is at $\langle \phi \rangle=0$, 
which gives 
\be 
\langle \Sigma \rangle = \langle g_\phi \rangle = \langle g \rangle  =0 \, .
\ee 
During inflation one can see that $m_{3/2} > H^2$, 
therefore these models are also within the perturbative unitarity bound \cite{DallAgata:2014qsj}. 
For the inflaton and gravitino masses in the vacuum we have 
\be
m_\phi|_\text{vacuum} \sim \sqrt{V_0} \ , \quad m_{3/2}|_\text{vacuum} = |g_0| \, , 
\ee
which realize \eqref{mmvac}. 
Therefore during preheating the longitudinal gravitino components produced are to be interpreted as  
inflatinos and stabilinos, 
instead of true vacuum goldstinos that might be cosmologically hazardous.

\subsection{Effective theories during the (p)reheating stage}

In this section we outline some basics of the gravitino production during (p)reheating in standard supergravity  
 considering the properties of the constrained superfield models discussed in the previous subsections. 
In the evolution of the universe subsequent to inflation, the inflaton field coherently oscillates about its vacuum and, among other particles,  longitudinal gravitinos get produced due to the non-adiabatical change of the frequency of the gravitino $\pm 1/2$ helicity modes. The resulting number density of the longitudinal gravitinos is found to be  \cite{Kallosh:1999jj, Giudice:1999fb, Giudice:1999yt, Giudice:1999am}
\begin{equation}
n_{l} (t)=\left\langle 0 | N/V| 0 \right\rangle =\frac{1}{\pi^2 a^3(t)} \int^{k_\text{max}}_0 dk\, k^2 |\beta_k|^2  \, , 
\end{equation} 
where $k_{\text{max}}$ is the comoving momentum cut-off of the effective theory and $\beta_k$ the corresponding Bogolyubov coefficient.  If the physical momentum cut-off is of the order of the inflaton mass scale or somewhat lower, the resulting longitudinal gravitino number density might be cosmologically problematic. Namely, if $n_l(t_\text{preh})\sim k^3_\text{max}/a^3(t_\text{preh}) \sim m^3_\phi$ the extrapolation of the longitudinal gravitino number density at the time of the inflaton perturbative decay, $t_\text{rh}$, gives the yield 
\begin{equation} \label{Ypre1}
\left. Y_l \right|_\text{preh}=\frac{n_l}{s} \simeq \frac{n_l(t_\text{preh})(t_\text{preh}/t_\text{rh})^2}{0.44 g_* T^3_\text{rh}} \, ,
\end{equation}
which rewrites
\begin{equation} \label{Ypre}
\left. Y_l \right|_\text{preh} \sim 10^{-14}\, \alpha^2\, \left(\frac{T_\text{rh}}{10^{10}\,\text{GeV}} \right) \, .
\end{equation}
 We considered that $a(t)\sim t^{2/3}$ during inflaton oscillations, $t_\text{preh}\sim \alpha m^{-1}_\phi$ with $\alpha\gg 1$ the characteristic time scale of preheating production,  $t^{-1}_\text{rh} \sim H_\text{rh} \sim g^{1/2}_* T^2_\text{rh} /M_\text{Pl}$ and $T_\text{rh}$ is the reheating temperature. If the above yield is interpreted as true vacuum goldstinos it generally violates the BBN bound (\ref{c2}) and/or might generate an overabundant dark matter at the time of the gravitino decay. Otherwise the estimation (\ref{Ypre1})-(\ref{Ypre}) is misleading. Evidently, the identification of the fermions that comprise the longitudinal gravitino component during the universe evolution is of central importance. To this end the field content of the theory and the corresponding dynamics should be specified.

As an illustrative example we choose a simple K\"ahler potential and superpotential for the inflaton--Polonyi  sector 
\be
\label{Wpreh}
K = X \overline X + \Phi \overline \Phi \ , \quad W = f_0 X + g_0 + \frac12 m \Phi^2 \, . 
\ee
The constrained superfields $X$ and $\Phi$ satisfy \eqref{X2} and \eqref{inflaton}, 
but we have rescaled the fermion and scalar components of $\Phi$ such that 
\be
\Phi = \frac{\phi + i b}{\sqrt 2} + \sqrt 2 \theta \chi + \theta^2 F^\Phi \, , 
\ee
where $b$ has the form \eqref{bEXPANSION} after appropriate rescaling of the component fields of $\Phi$. 
As explained in \cite{ Nilles:2001fg, Nilles:2001ry}, 
for models with 
K\"ahler potential and superpotential \eqref{Wpreh}, the longitudinal gravitino component $\psi^l$ is 
roughly given by 
\be
\label{long}
\psi^l \sim \sqrt{r_X} \, G  +  \sqrt{r_\Phi} \, \chi \, , 
\ee
where 
\be
\label{rxrf}
r_X = \frac{f_0^2}{f_0^2 + \frac12 \dot \phi^2 + \frac12 m^2 \phi^2} 
= \frac{\rho_X}{\rho_X+\rho_\Phi} \ ,  \quad 
r_\Phi =  \frac{\frac12 \dot \phi^2 + \frac12 m^2 \phi^2}{f_0^2 + \frac12 \dot \phi^2 + \frac12 m^2 \phi^2} 
= \frac{\rho_\Phi}{\rho_X+\rho_\Phi} \, . 
\ee 
From \eqref{long} and \eqref{rxrf} we see that if the inflaton superfield energy density $\rho_\Phi$ 
dominates over the Polonyi superfield energy density $\rho_X$ during preheating, 
then the inflaton oscillation parametrically excites longitudinal gravitinos  that should be identified as inflatinos \cite{Nilles:2001ry,Nilles:2001fg, Greene:2002ku}. 
In particular one has during preheating 
\be
\label{prehreq}
\rho_X|_\text{preh} \ll \rho_\Phi|_\text{preh} \ \rightarrow \ \psi^l |_\text{preh} \sim \chi \, , 
\ee
whereas at later times in the true vacuum, where the inflaton is strongly stabilized 
\be
\label{vacreq}
\rho_X|_\text{vacuum} \gg \rho_\Phi|_\text{vacuum} \ \rightarrow \ \psi^l |_\text{vacuum} \sim G \, . 
\ee
Numerical results can be found in the standard supergravity literature \cite{Nilles:2001ry,Nilles:2001fg, Greene:2002ku}. 
In models with the stabilizer superfield $S$, which has been studied recently \cite{Ema:2016oxl}, the 
description \eqref{long} for the longitudinal gravitino component 
will include an additional 
contribution given by $\sqrt{r_S} \, \lambda$ where $r_S = |F^S|^2/\rho_\text{total}$. These considerations imply that the yield (\ref{Ypre1})-(\ref{Ypre}) ought not be regarded as a realistic result.

In the working models studied in the previous subsections (4.1) and (4.2) one can easily check that the above results apply 
because the requirements \eqref{prehreq} and \eqref{vacreq} are met. 
In fact notice that  
if we expand the superpotentials from our previous examples in $\Phi$ (for $\Phi<M_\text{Pl}$) and keep only the leading terms 
we find the functions of the form \eqref{Wpreh}, 
and then follow Planck-suppressed ${\cal O}(\Phi/M_\text{Pl})$ terms.

Finally we comment on the perturbative decay of the inflaton to gravitino. 
Here the standard results apply (see e.g. \cite{Endo:2006zj, Nakamura:2006uc,Dine:2006ii,Kawasaki:2006hm}) 
and we have a small gravitino production due to the small value of $W_\Phi|$. 
In particular for the examples we presented in the previous subsections, 
we have $\langle g_\phi \rangle=0$, hence 
\be \label{Gt}
\Gamma(\phi \rightarrow \psi^l \psi^l) \sim \frac{|g_\phi|^2 m_\phi^5}{m_{3/2}^4} \ll \Gamma(\phi \rightarrow \text{MSSM})\,, 
\ee
that is the vacuum decay rate of the inflaton into two  gravitinos can be sufficiently suppressed compared to the total inflaton decay rate in order that nonthermal gravitino overproduction problems are avoided.

\section{Discussion}

The main motivation of our work was to show that there do exist single-field inflationary models 
in supergravity constructed from constrained superfields which have two properties: 
a) They contain a minimal bosonic sector and b) They reproduce the results from standard supergravity for (p)reheating processes.

To achieve this we have presented a minimal setup for constructing effective theories for inflation in supergravity, 
such that the (p)reheating phase does not directly lead to gravitino overproduction. 
We have introduced a new customized constrained superfield $\Phi$ 
which includes the inflaton but also its fermion superpartner and the auxiliary field. 
We presented simple inflationary models which utilize this constrained superfield. 
It is now very interesting to couple these models to matter and study how the reheating works 
in a realistic setup.

We believe that our work offers a strong motivation in favor of using constrained superfields 
for constructing {\it effective theories}, 
against the compensator method for implementing non-linear realizations. 
Indeed, the component form methods where the Volkov--Akulov goldstino is used as a compensator to implement the local non-linear 
realization of supersymmetry will in principle 
not contain neither the required superpartner for the inflaton nor the inflaton multiplet auxiliary field. 
Therefore, unless additional fields with very specific couplings are introduced, 
supersymmetry breaking comes always from the same sector, 
the Volkov--Akulov sector, 
implying that the gravitino yield at the course of the cosmological evolution might be wrongly overestimated.

\bigskip

\section*{Acknowledgments}

\noindent 
We thank N. Cribiori, G. Dall'Agata, A. Kehagias, A. Riotto and A. Van Proeyen. 
I.D. would like to thank the University of Padova for the hospitality offered to him during the preparation of this work. 
The work of I.D. is supported by the IKY Scholarship Programs for Strengthening Post Doctoral Research, co-financed by the European Social Fund ESF and the Greek government. The work of F.F. is supported in parts by the Padova University Project CPDA119349.

\end{document}